\documentclass[11pt,a4paper]{article}
\usepackage{jcappub}

\newcommand{\nn}{\nonumber \\}
\newcommand{\bea}{\begin{eqnarray}}
\newcommand{\ena}{\end{eqnarray}}

\newcommand{\RR}{{\cal R}}

\newcommand{\PP}{{\cal P}}

\title{Uncorrelated estimates of the primordial power spectrum}

\author{Zong-Kuan Guo}
\author{and Yuan-Zhong Zhang}

\affiliation{
State Key Laboratory of Theoretical Physics, Institute of Theoretical Physics,\\
Chinese Academy of Sciences, P.O. Box 2735, Beijing 100190, China}

\emailAdd{guozk@itp.ac.cn}
\emailAdd{zyz@itp.ac.cn}

\abstract{
We use the localized principle component analysis to
detect deviations from scale invariance of the primordial power
spectrum of curvature perturbations. With the technique we make
uncorrelated estimates of the primordial power spectrum with five
wavenumber bins.
In the framework of a minimal $\Lambda$CDM model,
using the latest cosmic microwave background data from the
WMAP and ACT experiments we find that
more than 95\% of the preferred models are incompatible with
the assumption of scale-invariance, but still compatible with a
power-law primordial spectrum.
We also forecast the sensitivity and constraints
achievable by the Planck experiment by performing Monte Carlo
studies on simulated data. Planck could significantly improve the
constraints on the primordial power spectrum, especially at small
scales by roughly a factor of 4.
}

\keywords{inflation, cosmological parameters from CMBR}

\begin{document}
\maketitle

\section{Introduction}

Measurements of anisotropies in the cosmic microwave background
(CMB) have played an essential role in constraining on basic
cosmological parameters, especially in probing the dynamics of
inflationary phase in the early Universe~\cite{tro02}.
The Wilkinson Microwave Anisotropy Probe (WMAP) satellite has
measured the CMB over the full sky down to $0.2^\circ$
resolution~\cite{kom08,kom10}.
Measurements at higher resolution made with the Atacama Cosmology
Telescope (ACT)~\cite{dun10} and the South Pole Telescope
(SPT)~\cite{kei11} can provide us with complementary
information about the early Universe on scales smaller than
those probed by the WMAP satellite.
The ACT experiment now measures fluctuations on scales from
$0.4^\circ$ to an arcminute.
A combination of the WMAP data and the ACT data would improve
the constraints on the cosmological parameters.

Inflationary models with featureless potentials generically
predict a primordial power spectrum of curvature perturbations
close to scale invariant.
Such models are usually parameterized by an amplitude of
spectrum $A_s$, a spectral index $n_s$ and its running
index $\alpha_s$ as
\bea
\ln \PP_\RR(k) = \ln A_s + (n_s-1)\ln(\frac{k}{k_0}) + \alpha_s \ln^2(\frac{k}{k_0})\;,
\ena
where $k_0$ is a pivot scale.
The parameterization is a Taylor expansion in the logarithmic
amplitude and logarithmic wavenumber space around the pivot point.
The special case with $n_s=1$ and $\alpha_s=0$ results in the
Harrison-Zel'dovich (scale invariant) spectrum.
In the slow-roll inflationary models, the spectral index and
running index are first and second order in the slow-roll
parameters respectively, and thus they are expected to be small.
For a power-law parameterization ($\alpha_s=0$), 99.5\% of
the preferred models are incompatible with the scale-invariant
spectrum by using the 7-year WMAP data
if tensor modes are ignored~\cite{kom10}.
Even adding a running index, a slightly tilted power-law primordial
power spectrum without tensor modes is still an excellent fit to
the data~\cite{kom10}.
Although, by combining the WMAP data with the ACT data,
the running index prefers a negative value at 1.8$\sigma$,
indicating enhanced damping at small scales,
there is no statistically significant deviation from a power-law
spectrum~\cite{dun10}.
Moreover, before claiming that a power-law spectrum is excluded,
one should investigate extensions of the minimal $\Lambda$CDM
model which could produce a similar effect in the CMB spectrum
(but not necessarily in the large-scale-structure power spectrum).
For instance, the marginal indication for enhanced damping in
the small scale CMB spectrum could also be explained by extra
relativistic degree of freedom~\cite{hou11}.
Other simple extensions of the minimal $\Lambda$CDM model include
small neutrino masses, a spatial curvature, a free primordial
helium fraction, etc. Here, we do not consider such alternatives,
stick to the minimal $\Lambda$CDM paradigm, and investigate only
the issue of the primordial spectrum beyond the power-law assumption.
Indeed, motivated by theoretical models or features of the observed data,
other various parameterizations of the primordial power spectrum
have been considered: for example, a broken power spectrum~\cite{bla03}
due perhaps to an interruption of the inflaton potential~\cite{bar01},
a cutoff at large scales~\cite{efs03,bri06}
motivated by suppression of the lower multipoles in the CMB anisotropies~\cite{nol08,cop10},
and more complicated shapes of the spectrum caused by
features in the inflaton potential~\cite{sta92}.

Measuring deviations from scale invariance of the primordial
power spectrum is a critical test of cosmological inflation.
Either exact scale invariance or a strong deviation from scale
invariance could falsify the idea of inflation.
However, a strong theory prior on the form of the primordial
power spectrum could lead to misinterpretation and biases in
parameter determination.
Some more general approaches have been proposed to reconstruct
the shape of the primordial power spectrum from existing data,
based on linear interpolation~\cite{bri03},
cubic spline interpolation in log-log space~\cite{guo11},
a minimally-parametric reconstruction~\cite{sea05},
wavelet expansions~\cite{muk03}, principle component analysis~\cite{hu03},
and a direct reconstruction via deconvolution methods~\cite{kog03,sha03,toc04}.
The first three approaches are sensitive to the overall shape
of the spectrum while the last three reconstruct methods are
sensitive to the local features in the spectrum.
Therefore they are complementary and needed to cross-check each other.

In this work we focus on the uncorrelated band-power estimates
of the primordial power spectrum to measure deviations from scale
invariance, based on the local principle component analysis
introduced to study dark energy in~\cite{hut04}.
We apply the method to the 7-year WMAP data~\cite{kom10} and
in combination with small-scale CMB data from the ACT experiment~\cite{dun10}.
In our analysis we adopt two main astrophysical priors on the
Hubble constant ($H_0$) measured from the magnitude-redshift
relation of 240 low-$z$ Type Ia supernovae at $z<0.1$~\cite{rie09}
and on the distance ratios of the comoving sound horizon to
the angular diameter distances from the Baryon Acoustic
Oscillation (BAO) in the distribution of galaxies~\cite{per09}.
Moreover, we generate mock data for the Planck experiment and then
make forecast using the Monte Carlo simulation approach.
As expected, Planck could significantly improve the constraints
on the primordial power spectrum.

This paper is organized as follows. In Section~\ref{sec2},
we first describe the method and the data used in this analysis.
We then apply the method to the seven-year WMAP data and
in combination with the ACT data and present our
results.
In Section~\ref{sec3}, using a Monte Carlo approach we analyze
the sensitivity of the Planck experiment with respect to
the primordial power spectrum.
Section~\ref{sec4} is devoted to conclusions.

\section{Uncorrelated constraints from current observations}
\label{sec2}

We consider a spatially flat $\Lambda$CDM Universe described by
the following cosmological parameters
\bea
\left\{\Omega_b h^2, \Omega_c h^2, \Theta_s, \tau, A_1, A_2, ..., A_5 \right\},
\ena
where $\Omega_b h^2$ and $\Omega_c h^2$ are the physical baryon and
cold dark matter densities relative to the critical density,
$h$ is the dimensionless Hubble parameter such that $H_0=100h$
kms$^{-1}$Mpc$^{-1}$,
$\Theta_s$ is the ratio of the sound horizon to the angular diameter
distance at decoupling, and $\tau$ is the reionization optical depth.
Since we do not consider extensions of the minimal flat $\Lambda$CDM
model in this analysis, we fixed the primordial helium fraction
and effective neutrino number to their standard values, and did not
introduce neutrino masses or a tensor modes.
The primordial power spectrum parameters,
$A_i \equiv \ln \left[10^{10} \PP_{\RR}(k_i)\right]$ ($i=1,2,...,5$),
are the logarithmic values of the primordial power spectrum
of curvature perturbations $\PP_{\RR}(k)$
at five knots $k_i$, equally spaced in logarithmic wavenumber between
$0.0002$ Mpc$^{-1}$ and $0.2$ Mpc$^{-1}$.
To reconstruct a smooth spectrum with continuous first and second
derivatives with respect to $\ln k$, we use a cubic spline interpolation
to determine logarithmic values of the primordial power spectrum
between these nodes.
Outside of the wavenumber range we fix the slope of the primordial
power spectrum at the boundaries since the CMB data place only weak
constraints on them.
Using $\ln \PP_{\RR}(k)$ instead of $\PP_{\RR}(k)$ for splines
ensures the positive definiteness of the primordial power
spectrum at the expense of making the primordial power spectrum
non-linear in the parameters.
Otherwise, we must discard such steps in the Markov chain if the
interpolating spline between the knots goes negative due to steep
slopes~\cite{sea05}.
As discussed in~\cite{guo11}, the method is insensitive to local
features in the primordial power spectrum, but is sensitive to
the overall shape.

The primordial spectrum parameters, $A_i$, are correlated due
to the geometric projection from the primordial power spectrum to
the angular power spectrum and gravitational lensing.
These correlations are encapsulated in the covariance matrix of
the primordial spectrum parameters,
\bea
{\bf C}=(A_i-\langle A_i \rangle)(A_j-\langle A_j \rangle)^{T},
\ena
which can be obtained by taking the average of the Markov chain
and marginalizing over other cosmological parameters.
The diagonal elements of the covariance matrix are the variances
of $A_i$ and the non-diagonal elements represent corrections
between the $A_i$ bins that slowly decrease with increasing
bin separation.
To eliminate these correlations, we employ the localized principle
component analysis to construct a new basis,
where the new parameters $\tilde{A_i}$ are uncorrelated~\cite{ham99}.
This variant of the principal component analysis has recently been
applied to probe the dynamics of dark energy~\cite{hut04,zha08,ser09}.
We diagonalize the Fisher matrix ${\bf F} \equiv {\bf C}^{-1}$,
so that
\bea
{\bf F}={\bf O}^{T}{\bf D}{\bf O}\;,
\ena
where ${\bf O}$
is an orthogonal matrix and ${\bf D}$ is the diagonalized inverse
covariance of the transformed bins.
The localized principle component analysis corresponds to
the weight matrix ${\bf W}={\bf O}^{T}{\bf D}^{1/2}{\bf O}$,
which is usually normalized so that its rows sum up to unity.
The weights are fairly localized in wavenumber since ${\bf D}^{1/2}$
is absorbed into {\bf O}.
With this choice, the uncorrelated parameters can be now
obtained by changing the basis through the weight matrix rotation,
$\tilde{\bf A}={\bf W}{\bf A}$. When discussing our results,
we will generally refer to these uncorrelated estimates.

{\it Data:}
We use the 7-year WMAP data (WMAP7) and in combination with the
148 GHz ACT data during its 2008 season.
For the WMAP data, we use the low-$l$ and high-$l$ temperature
and polarization power spectra.
We also consider the Sunyaev-Zel'dovich (SZ) effect, in which
CMB photons scatter off hot electrons in clusters of galaxies.
Given a SZ template it is described by a SZ template amplitude
$A_{\rm SZ}$ as in the WMAP papers~\cite{kom08,kom10}.
For the ACT data, we focus on the band powers in the multiple
range $1000 \le l \le 3000$.
Following Ref.~\cite{dun10} for computational efficiency the
CMB is set to zero above $l=4000$ where the contribution is
subdominant, less than 5\% of the total power.
To use the ACT likelihood described in~\cite{dun10}, aside from
$A_{\rm SZ}$ there are two more secondary parameters, $A_p$ and $A_c$.
The former is the total Poisson power at $l=3000$ from radio and
infrared point sources. The latter is the template amplitude of
the clustered power from infrared point sources.
We impose positivity priors on the three secondary parameters,
use the SZ template and the clustered source template provided by
the ACT likelihood package, and marginalize over these secondary
parameters to account for SZ and point source contamination.
We adopt two main astrophysical priors: the present-day
Hubble constant $H_0=74.2\pm 3.6$ km s$^{-1}$ Mpc$^{-1}$
measured from the magnitude-redshift relation of
240 low-$z$ Type Ia supernovae at $z<0.1$~\cite{rie09}, and
the distances ratios, $r_s/D_V(z=0.2)=0.1905 \pm 0.0061$ and
$r_s/D_V(z=0.35)=0.1097 \pm 0.0036$,
measured from the two-degree field galaxy redshift survey and
the sloan digital sky survey data~\cite{per09}.
Here $r_s$ is the comoving sound horizon size at the baryon
drag epoch and $D_V$ is the effective distance measure for
angular diameter distance.

\begin{figure}[!htb]
\begin{center}
\includegraphics[width=70mm]{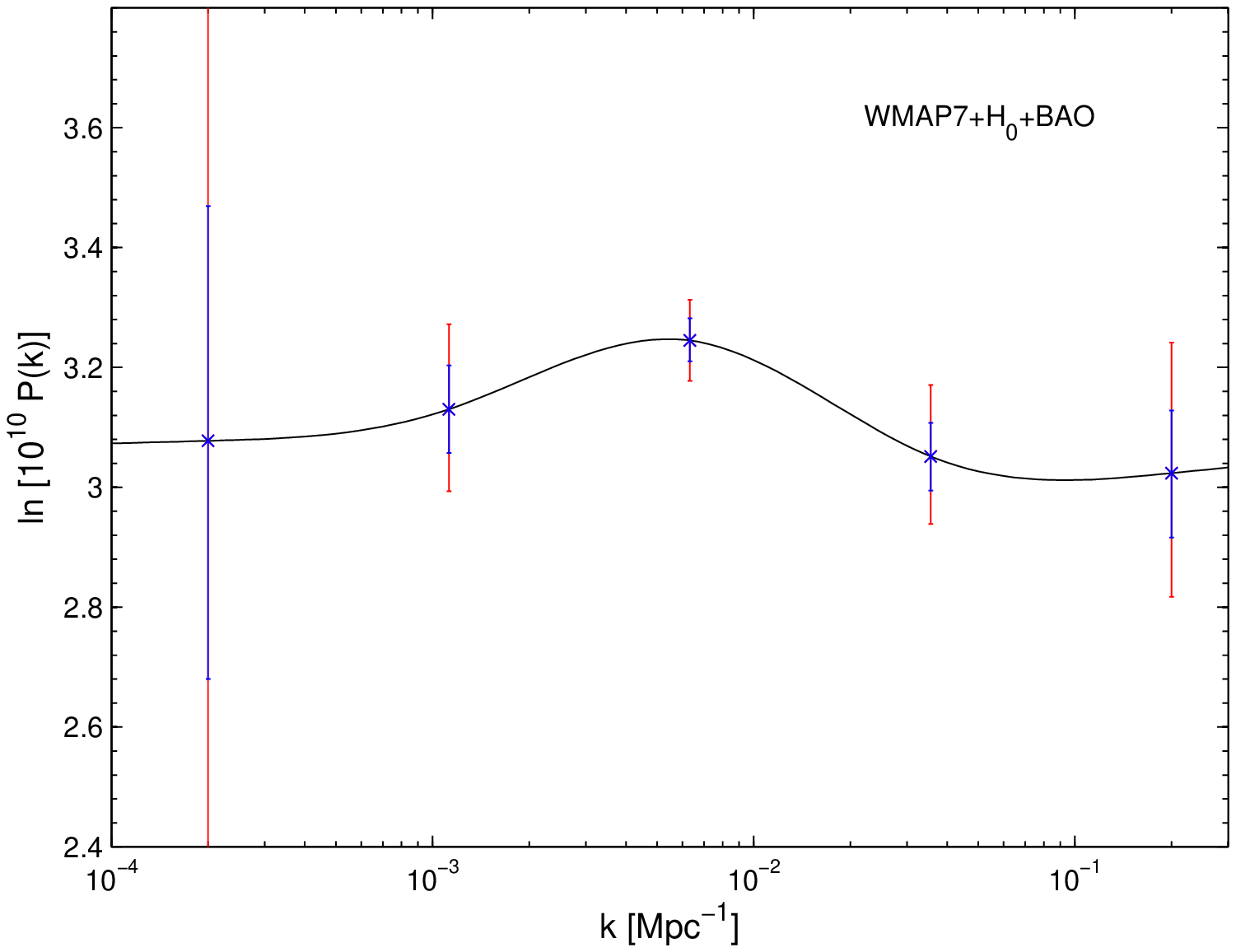}
\includegraphics[width=70mm]{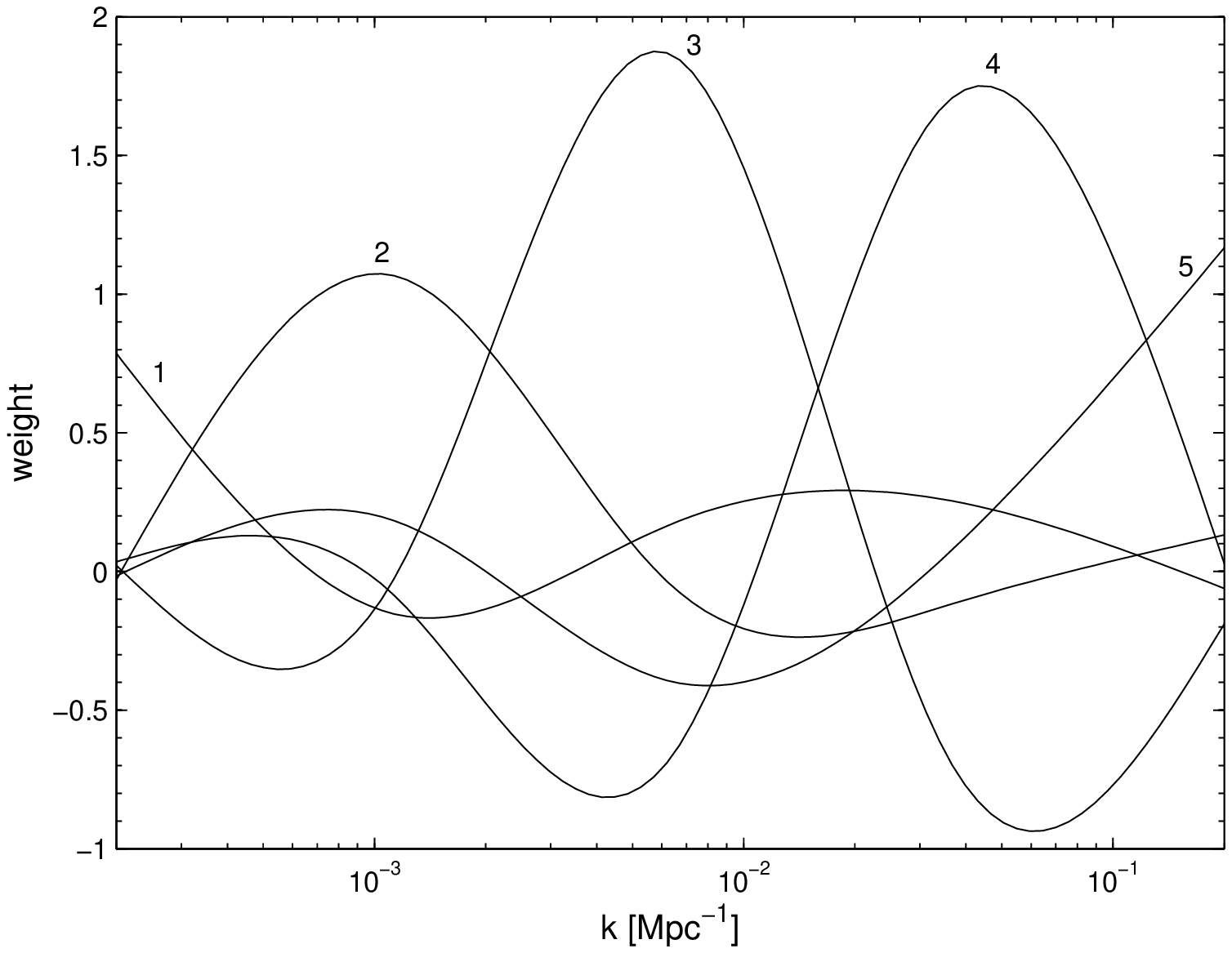}
\includegraphics[width=70mm]{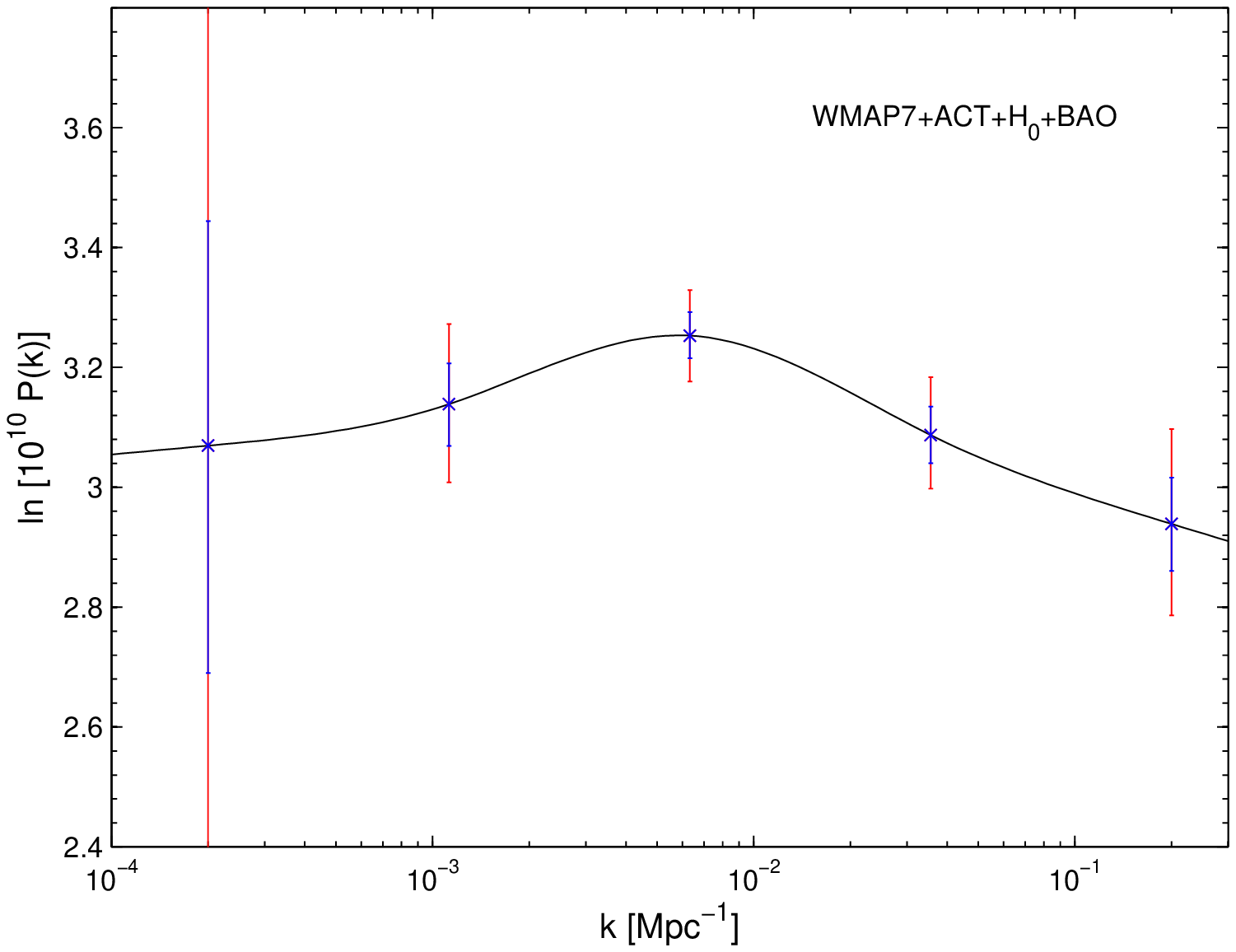}
\includegraphics[width=70mm]{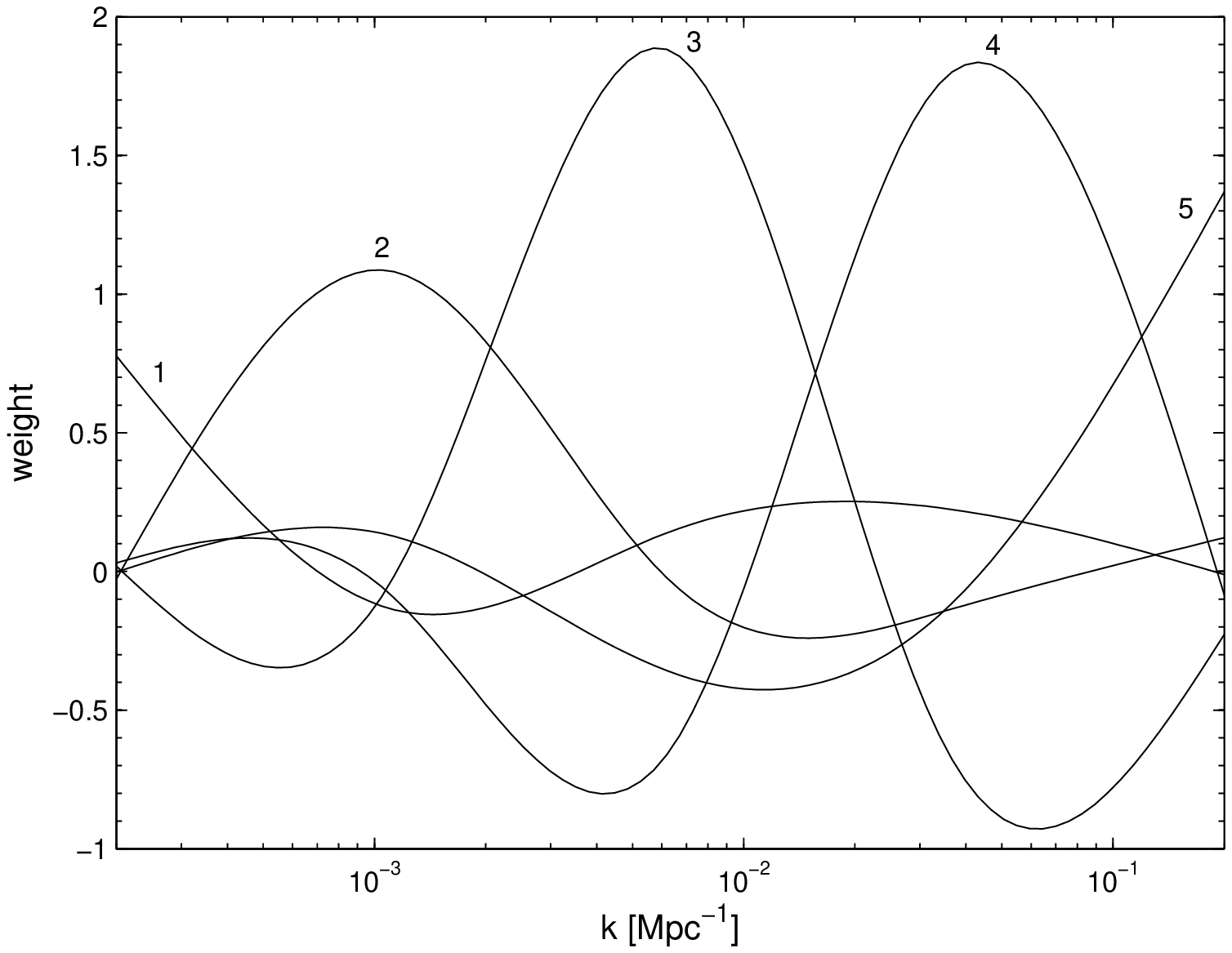}
\caption{Uncorrelated constraints on the primordial power spectrum
of curvature perturbations and their weight functions,
derived from the WMAP7+$H_0$+BAO combination (top panels) and
from the WMAP7+ACT+$H_0$+BAO combination (bottom panels).
The blue and red error bars show $1\sigma$ and $2\sigma$ uncertainties
respectively.}
\label{figwmap}
\end{center}
\end{figure}

{\it Results:}
Our analysis is carried out using a modified version of the
publicly available CosmoMC package, which explores the parameter space by
means of Monte Carlo Markov Chains~\cite{lew02}.
Figure~\ref{figwmap} shows the uncorrelated constraints on the
primordial power spectrum of curvature perturbations
(68\% and 95\% CL) and the corresponding
weight functions that describe transformation from correlated
parameters $A_i$ to the uncorrelated $\tilde{A}_i$,
derived from WMAP7+$H_0$+BAO (top panels)
and from WMAP7+ACT+$H_0$+BAO (bottom panels), respectively.
We can see that the power spectrum is best determined
around $k \sim 0.007$ Mpc$^{-1}$, and less accurately determined
at much lower and much higher wavenumber because of the cosmic
variance and dominant noise respectively.
As shown in the top-left panel of figure~\ref{figwmap},
95\% of the preferred models are incompatible with the
assumption of scale-invariance, but still compatible with a
power-law primordial spectrum.
Adding the ACT data we find that there is more deviation
from a simple scale-invariant spectrum due to reduced errors
and suppressed spectrum at high-$k$, but it is weaker than
the corresponding result from WMAP adopting a inflation-motivated
power-law spectrum prior~\cite{kom10}.
Note that the weights are fairly localized in $k$,
as found in the context of dark energy measurements~\cite{hut04,zha08,ser09}.
Moreover, the weight functions in the top-right panel are
similar to those in the bottom-right panel.

\begin{table}[!htb]
\begin{center}
\begin{tabular}{lcccccc}
\hline
Parameter & WMAP7+$H_0$+BAO & WMAP7+ACT+$H_0$+BAO & Planck \\
\hline
$\tilde{A}_1$ & $3.0776\pm0.3973$ & $3.0696\pm0.3855$ & $3.1488\pm0.3212$ \\
$\tilde{A}_2$ & $3.1302\pm0.0720$ & $3.1388\pm0.0681$ & $3.1289\pm0.0379$ \\
$\tilde{A}_3$ & $3.2451\pm0.0356$ & $3.2532\pm0.0393$ & $3.1327\pm0.0243$ \\
$\tilde{A}_4$ & $3.0515\pm0.0584$ & $3.0872\pm0.0473$ & $3.1366\pm0.0122$ \\
$\tilde{A}_5$ & $3.0237\pm0.1076$ & $2.9390\pm0.0784$ & $3.1384\pm0.0167$ \\
\hline
\end{tabular}
\end{center}
\caption{Uncorrelated constraints on the primordial power spectrum
with 68\% confidence levels.}
\label{tabunc}
\end{table}

\section{Planck forecast constraints}\label{sec3}

In this section, we apply Monte Carlo Markov Chain methods
to assess the accuracy with which the primordial power spectrum
can be constrained from Planck experiment.
Following the approach described in Ref.~\cite{per06},
we generate synthetic data for the Planck experiment and then
perform a systematic analysis on the simulated data.
Assuming a fiducial $\Lambda$CDM model with a scale-invariant
power spectrum, one can use a Boltzmann code such as CAMB~\cite{lew99}
to calculate the angular power spectra $C_{l}^{TT}$, $C_{l}^{TE}$,
$C_{l}^{EE}$, $C_{l}^{dd}$ and $C_{l}^{Td}$ for the temperature,
cross temperature-polarization, polarization, deflection field
and cross temperature-defection.
We assume that beam uncertainties are small and that
uncertainties due to foreground removal are smaller than
statistical errors.
For an experiment with some known beam width and detectors
sensitivity, the noise power spectrum $N_l^{TT}$, $N_l^{EE}$
and $N_l^{dd}$ can be estimated.
Here we use the FuturCMB package\footnote{The FuturCMB package is available at: http://lpsc.in2p3.fr/perotto/}
to calculate $N_l^{dd}$ based on the quadratic estimator method proposed in~\cite{oka03},
which provides an algorithm for estimating the noise spectrum
of the deflection field from the observed CMB primary
anisotropy and noise power spectra.
For Planck we combine only the 100, 143 and 217 GHz HFI channels,
with beam width $\theta_{\rm FWHM}$ = ($9.6'$, $7.0'$, $4.6'$)
in arcminutes, temperature noise per pixel $\sigma_T$ = (8.2, 6.0, 13.1)
in $\mu K$ and polarization noise per pixel $\sigma_{E}$
= (13.1, 11.2, 24.5) in $\mu K$ (see Ref.~\cite{tau10} for the
instrumental specifications of Planck).
Given the fiducial spectra $C_l$ and noise spectra $N_l$, one can
generate mock data $\hat{C}_l$.
We perform a Monte Carlo analysis through the likelihood
function defined as
\bea
-2\ln {\cal L} = \sum_l (2l+1) f_{\rm sky} \left(\frac{D}{|\bar{C}|}
 + \ln \frac{|\bar{C}|}{|\hat{C}|} - 3\right),
\ena
where
\bea
D &=& \hat{C}_l^{TT}\bar{C}_l^{EE}\bar{C}_l^{dd}
 + \bar{C}_l^{TT}\hat{C}_l^{EE}\bar{C}_l^{dd}
 + \bar{C}_l^{TT}\bar{C}_l^{EE}\hat{C}_l^{dd} \nn
&& - \bar{C}_l^{TE}\left(\bar{C}_l^{TE}\hat{C}_l^{dd} + 2\hat{C}_l^{TE}\bar{C}_l^{dd}\right)
 - \bar{C}_l^{Td}\left(\bar{C}_l^{Td}\hat{C}_l^{EE} + 2\hat{C}_l^{Td}\bar{C}_l^{EE}\right), \\
|\bar{C}| &=& \bar{C}_l^{TT}\bar{C}_l^{EE}\bar{C}_l^{dd}
 -\left(\bar{C}_l^{TE}\right)^2 \bar{C}_l^{dd}-\left(\bar{C}_l^{Td}\right)^2 \bar{C}_l^{EE}, \\
|\hat{C}| &=& \hat{C}_l^{TT}\hat{C}_l^{EE}\hat{C}_l^{dd}
 -\left(\hat{C}_l^{TE}\right)^2 \hat{C}_l^{dd}-\left(\hat{C}_l^{Td}\right)^2 \hat{C}_l^{EE}.
\ena
Here, $\bar{C}_l=C_l+N_l$ is the theoretical spectrum plus noise and
$f_{\rm sky}$ is the sky fraction due to foregrounds removal.
For Planck we choose $f_{\rm sky}$ = 0.65, corresponding to
a $\pm 20^\circ$ galactic cut, and consider the data up to $l=2000$.

\begin{figure}[!htb]
\begin{center}
\includegraphics[width=70mm]{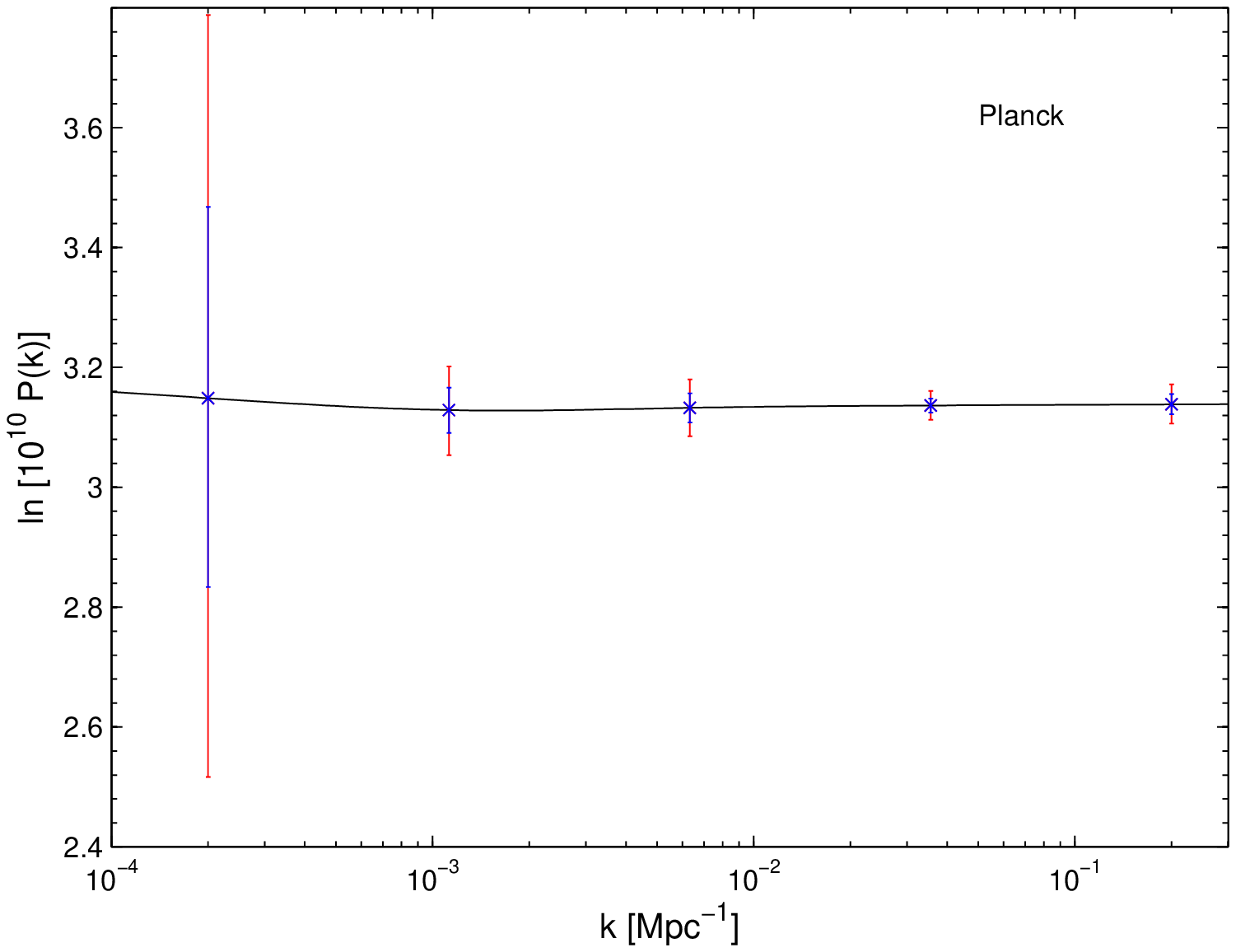}
\includegraphics[width=70mm]{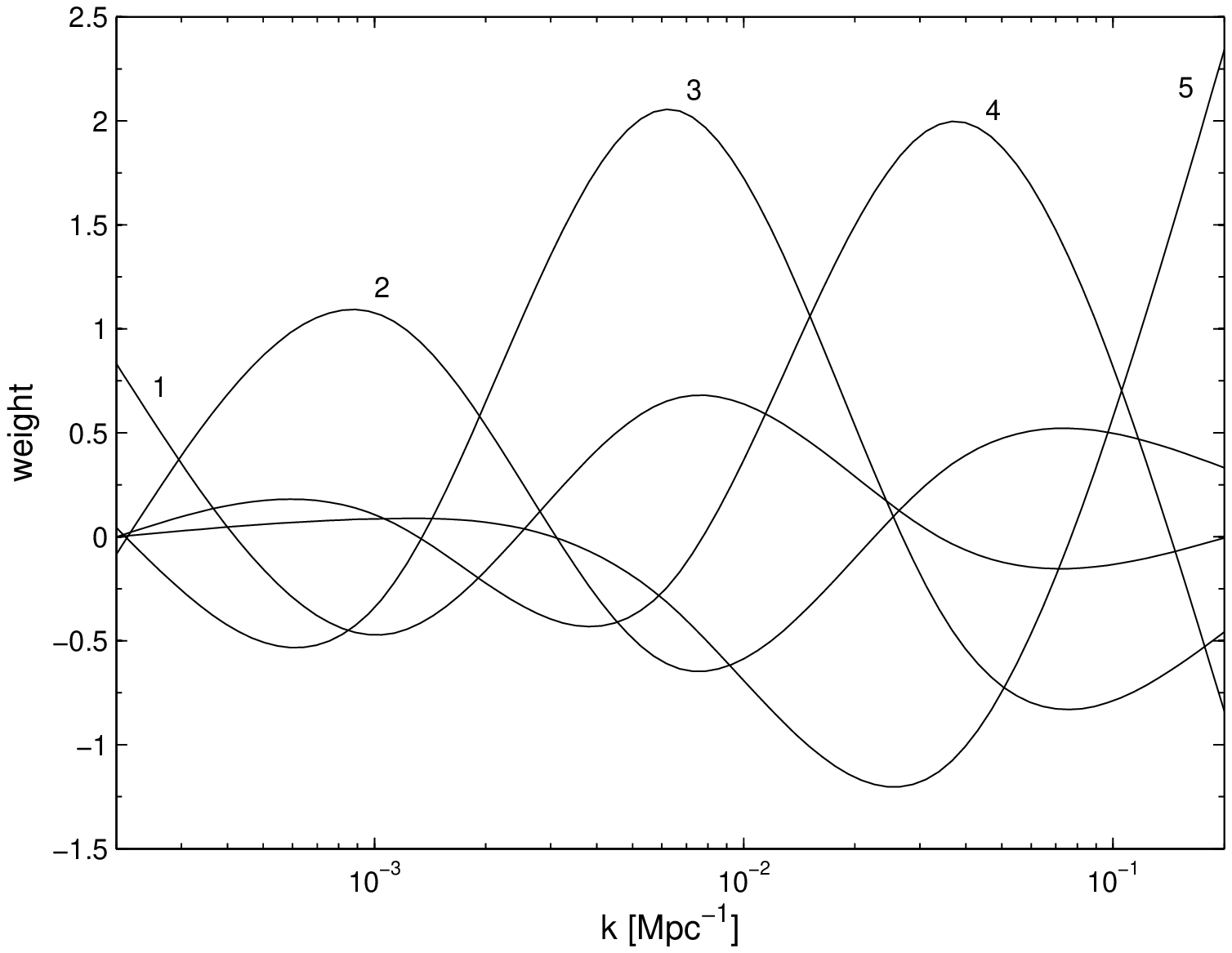}
\caption{Uncorrelated constraints on the primordial power spectrum
of curvature perturbations and their weight functions
from Planck simulated data.
The blue and red error bars show $1\sigma$ and $2\sigma$ uncertainties
respectively.}
\label{figplanck}
\end{center}
\end{figure}

Our results are presented in figure~\ref{figplanck} and table~\ref{tabunc}
for Planck simulated data.
As we can see in table~\ref{tabunc}, Planck will reduce the
uncertainties in $\tilde{A}_i$, especially in $\tilde{A}_4$ by
a factor of 3.9 and in $\tilde{A}_5$ by a factor of 4.7.
Since large uncertainties of the power spectrum at low-$k$ mainly
arise from the cosmic variance, measurement of $\tilde {A}_1$ is limited.
The weight functions in the right panel of figure~\ref{figplanck}
are a little better localized in wavenumber than those in
figure~\ref{figwmap} because the weak lensing effect
is extracted from CMB maps provided by Planck.
Furthermore, we have checked that the weight functions depend
weakly on the fiducial cosmological model.

\section{Conclusions}\label{sec4}

Most inflationary models predict small deviations from a
scale-invariant power spectrum.
Therefore, the measurements of deviations from an exact scale-invariant
spectrum would provide a firm probe of the dynamics of an
inflationary phase happened in the early Universe.
The local principal component technique is a powerful tool for measuring
deviations from the scale-invariant spectrum, complementary to
other approaches to reconstruct the primordial power spectrum
or the direct testing of slow-roll inflation~\cite{mar03,cli06,les07}.
In this paper, we have used the localized principal component
analysis to produce uncorrelated estimates of the primordial
power spectrum of curvature perturbations.
In the framework of a minimal $\Lambda$CDM model, we found
that more than 95\% of the preferred models are incompatible
with the scale-invariant spectrum, but still compatible with a
power-law primordial spectrum by using the 7-year WMAP data
in combination with the ACT data.
This conclusion is a little stronger than the corresponding result
in Ref.~\cite{guo11}, but weaker than when the inflation-motivated
power-law prior is adopted.
We have performed a systematic analysis of the future constraints
on the primordial power spectrum achievable from the Planck experiment.
We found that Planck would be able to shrink the error bars on
the spectrum bins especially at small scales by roughly a factor of 4,
which is promising to definitively detect these deviations.

\acknowledgments
Our numerical analysis was performed on the Lenovo DeepComp 7000 supercomputer in SCCAS.
This work is partially supported by the project of Knowledge Innovation
Program of Chinese Academy of Science and
National Basic Research Program of China under Grant No:2010CB832805.
We used CosmoMC and CAMB.
We also acknowledge the use of the Legacy Archive for Microwave Background Data Analysis and ACT data.


\begin{thebibliography}{99}
\bibitem{tro02}
R.~Trotta, A.~Riazuelo and R.~Durrer,
  Phys.\ Rev.\ D {\bf 67}, 063520 (2003)
  [arXiv:astro-ph/0211600];
R.~Trotta,
  New\ Astron.\ Rev.\ {\bf 47}, 769 (2003)
  [arXiv:astro-ph/0304525];
R.~Trotta,
  arXiv:astro-ph/0410115;
R.~Trotta,
  Contemp.\ Phys.\ {\bf 49}, 71 (2008)
  [arXiv:0803.4089].
\bibitem{kom08}
E.~Komatsu, {\it et al.},
  Astrophys.\ J.\ Suppl.\ {\bf 180}, 330 (2009)
  [arXiv:0803.0547].
\bibitem{kom10}
E.~Komatsu, {\it et al.},
  Astrophys.\ J.\ Suppl.\ {\bf 192}, 18 (2011)
  [arXiv:1001.4538].
\bibitem{dun10}
J.~Dunkley, {\it et al.},
  arXiv:1009.0866;
R.~Hlozek, {\it et al.},
  arXiv:1105.4887.
\bibitem{kei11}
R.~Keisler, {\it et al.},
  arXiv:1105.3182.
\bibitem{hou11}
Z.~Hou, R.~Keisler, L.~Knox, M.~Millea and C.~Reichardt,
  arXiv:1104.2333.
\bibitem{bla03}
A.~Blanchard, M.~Douspis, M.~Rowan-Robinson and S.~Sarkar,
  Astron.\ Astrophys.\ {\bf 412}, 35 (2003)
  [arXiv:astro-ph/0304237].
\bibitem{bar01}
J.~Barriga, E.~Gaztanaga, M.~Santos and S.~Sarkar
  Mon.\ Not.\ Roy.\ Astron.\ Soc.\ {\bf 324}, 977 (2001)
  [arXiv:astro-ph/0011398].
\bibitem{efs03}
G.~Efstathiou,
  Mon.\ Not.\ Roy.\ Astron.\ Soc.\ {\bf 343}, L95 (2003)
  [arXiv:astro-ph/0303127].
\bibitem{bri06}
M.~Bridges, A.~N.~Lasenby and M.~P.~Hobson,
  Mon.\ Not.\ Roy.\ Astron.\ Soc.\ {\bf 369}, 1123 (2006)
  [arXiv:astro-ph/0511573];
M.~Bridges, A.~N.~Lasenby and M.~P.~Hobson,
  Mon.\ Not.\ Roy.\ Astron.\ Soc.\ {\bf 381}, 68 (2007)
  [arXiv:astro-ph/0607404].
\bibitem{nol08}
M.~R.~Nolta, {\it et al.},
 Astrophys.\ J.\ Suppl.\ {\bf 180}, 296 (2009)
 [arXiv:0803.0593].
\bibitem{cop10}
C.~J.~Copi, D.~Huterer, D.~J.~Schwarz and G.~D.~Starkman,
  Adv.\ Astron.\ {\bf 2010}, 847541 (2010)
  [arXiv:1004.5602].
\bibitem{sta92}
A.~A.~Starobinsky,
  JETP Lett. {\bf 55}, 489 (1992);
J.~A.~Adams, G.~G.~Ross and S.~Sarkar,
  Nucl.\ Phys.\ B {\bf 503}, 405 (1997)
  [arXiv:hep-ph/9704286].
\bibitem{bri03}
S.~L.~Bridle, A.~M.~Lewis, J.~Weller and G.~Efstathiou
  Mon.\ Not.\ Roy.\ Astron.\ Soc.\ {\bf 342}, L72 (2003)
  [arXiv:astro-ph/0302306];
S.~Hannestad,
  JCAP {\bf 0404}, 002 (2004)
  [arXiv:astro-ph/0311491].
\bibitem{guo11}
Z.~K.~Guo, D.~J.~Schwarz and Y.~Z.~Zhang,
  JCAP {\bf 08}, 031 (2011)
  [arXiv:1105.5916].
\bibitem{sea05}
C.~Sealfon, L.~Verde and R.~Jimenez,
  Phys.\ Rev.\ D {\bf 72}, 103520 (2005)
  [arXiv:astro-ph/0506707];
L.~Verde and H.~V.~Peiris,
  JCAP {\bf 0807}, 009 (2008)
  [arXiv:0802.1219];
H.~V.~Peiris and L.~Verde,
  Phys.\ Rev.\ D {\bf 81}, 021302 (2010)
  [arXiv:0912.0268].
\bibitem{muk03}
P.~Mukherjee and Y.~Wang,
  Astrophys.\ J.\ {\bf 598}, 779 (2003)
  [arXiv:astro-ph/0301562 ];
P.~Mukherjee and Y.~Wang,
  Astrophys.\ J.\ {\bf 599}, 1 (2003)
  [arXiv:astro-ph/0303211];
P.~Mukherjee and Y.~Wang,
  JCAP {\bf 0512}, 007 (2005)
  [arXiv:astro-ph/0502136].
\bibitem{hu03}
W.~Hu and T.~Okamoto,
  Phys.\ Rev.\ D {\bf 69}, 043004 (2004)
  [arXiv:astro-ph/0308049];
S.~Leach,
  Mon.\ Not.\ Roy.\ Astron.\ Soc.\ {\bf 372}, 646 (2006)
  [arXiv:astro-ph/0506390].
\bibitem{kog03}
N.~Kogo, M.~Matsumiya, M.~Sasaki and J.~Yokoyama,
  Astrophys.\ J.\ {\bf 607}, 32 (2004)
  [arXiv:astro-ph/0309662];
R.~Nagata and J.~Yokoyama,
  Phys.\ Rev.\ D {\bf 78}, 123002 (2008)
  [arXiv:0809.4537];
R.~Nagata and J.~Yokoyama,
  Phys.\ Rev.\ D {\bf 79}, 043010 (2009)
  [arXiv:0812.4585];
K.~Ichiki and R.~Nagata,
  Phys.\ Rev.\ D {\bf 80}, 083002 (2009);
K.~Ichiki, R.~Nagata and J.~Yokoyama,
  Phys.\ Rev.\ D {\bf 81}, 083010 (2010)
  [arXiv:0911.5108].
\bibitem{sha03}
A.~Shafieloo and T.~Souradeep,
  Phys.\ Rev.\ D {\bf 70}, 043523 (2004)
  [arXiv:astro-ph/0312174];
A.~Shafieloo and T.~Souradeep,
  Phys.\ Rev.\ D {\bf 78}, 023511 (2008)
  [arXiv:0709.1944];
J.~Hamann, A.~Shafieloo and T.~Souradeep,
  JCAP {\bf 1004}, 010 (2010)
  [arXiv:0912.2728].
\bibitem{toc04}
D.~Tocchini-Valentini, M.~Douspis and J.~Silk,
  Mon.\ Not.\ Roy.\ Astron.\ Soc.\ {\bf 359}, 31 (2005)
  [arXiv:astro-ph/0402583];
D.~Tocchini-Valentini, Y.~Hoffman and J.~Silk,
  Mon.\ Not.\ Roy.\ Astron.\ Soc.\ {\bf 367}, 1095 (2006)
  [arXiv:astro-ph/0509478].
\bibitem{hut04}
D.~Huterer and A.~Cooray,
  Phys.\ Rev.\ D {\bf 71}, 023506 (2005)
  [arXiv:astro-ph/0404062];
D.~Sarkar, {\it et al.},
  Phys.\ Rev.\ Lett.\ {\bf 100}, 241302 (2008)
  [arXiv:0709.1150].
\bibitem{rie09}
A.~G.~Riess, {\it et al.},
  Astrophys.\ J.\ {\bf 699}, 539 (2009)
  [arXiv:0905.0695].
\bibitem{per09}
W.~J.~Percival, {\it et al.},
  Mon.\ Not.\ Roy.\ Astron.\ Soc.\ {\bf 401}, 2148 (2010)
  [arXiv:0907.1660].
\bibitem{ham99}
A.~J.~S.~Hamilton and M.~Tegmark,
  Mon.\ Not.\ Roy.\ Astron.\ Soc.\ {\bf 312}, 285 (2000)
  [arXiv:astro-ph/9905192].
\bibitem{zha08}
G.~B.~Zhao, D.~Huterer and X.~Zhang,
  Phys.\ Rev.\ D {\bf 77}, 121302 (2008)
  [arXiv:0712.2277];
G.~B.~Zhao and X.~Zhang,
  Phys.\ Rev.\ D {\bf 81}, 043518 (2010)
  [arXiv:0908.1568].
\bibitem{ser09}
P.~Serra, {\it et al.},
  Phys.\ Rev.\ D {\bf 80}, 121302 (2009)
  [arXiv:0908.3186].
\bibitem{lew02}
A.~Lewis and S.~Bridle,
  Phys.\ Rev.\ D {\bf 66}, 103511 (2002)
  [arXiv:astro-ph/0205436].
\bibitem{per06}
L.~Perotto, {\it et al.},
  JCAP {\bf 0610}, 013 (2006)
  [arXiv:astro-ph/0606227].
\bibitem{lew99}
A.~Lewis, A.~Challinor and A.~Lasenby,
  Astrophys.\ J.\ {\bf 538}, 473 (2000)
  [arXiv:astro-ph/9911177].
\bibitem{oka03}
T.~Okamoto and W.~Hu,
  Phys.\ Rev.\ D {\bf 67}, 083002 (2003)
  [arXiv:astro-ph/0301031].
\bibitem{tau10}
J.~A.~Tauber, {\it et al.},
  A\&A {\bf 520}, A1 (2010).
\bibitem{mar03}
J.~Martin and C.~Ringeval,
  Phys.\ Rev.\ D {\bf 69}, 083515 (2004)
  [arXiv:astro-ph/0310382];
J.~Martin and C.~Ringeval,
  Phys.\ Rev.\ D {\bf 69}, 127303 (2004)
  [arXiv:astro-ph/0402609];
J.~Martin and C.~Ringeval,
  JCAP {\bf 0501}, 007 (2005)
  [arXiv:hep-ph/0405249].
\bibitem{cli06}
J.~M.~Cline and L.~Hoi,
  JCAP {\bf 0606}, 007 (2006)
  [arXiv:astro-ph/0603403];
D.~K.~Hazra, {\it et al.},
  JCAP {\bf 1010}, 008 (2010)
  [arXiv:1005.2175];
M.~Aich, {\it et al.},
  arXiv:1106.2798.
\bibitem{les07}
J.~Lesgourgues and W.~Valkenburg,
  Phys.\ Rev.\ D {\bf 75}, 123519 (2007)
  [arXiv:astro-ph/0703625];
J.~Lesgourgues, A.~A.~Starobinsky and W.~Valkenburg,
  JCAP {\bf 0801}, 010 (2008)
  [arXiv:0710.1630];
F.~Finelli, J.~Hamann, S.~M.~Leach and J.~Lesgourgues,
  JCAP {\bf 1004}, 011 (2010)
  [arXiv:0912.0522].
\end{thebibliography}
\end{document}